# Bulk Photovoltaic Effect in Two-Dimensional Distorted MoTe$_2$

Sikandar Aftab [1], Muhammad Arslan Shehzad[2], Muhammad Salman Ajmal[3], Fahmid Kabir[4], and

Muhammad Zahir Iqbal[5]

[1] Department of Intelligent Mechatronics Engineering, Sejong University, Seoul 05006, South Korea. [2]Northwestern University Atomic and Nanoscale Characterization Experimental (NUANCE) Center, Northwestern University, Evanston, Illinois 60208, United States. [3]Department of Nanotechnology and Advanced Materials Engineering, Sejong University, Seoul 05006, South Korea. [4]School of Engineering Science, Simon Fraser University, Burnaby, British Columbia V5A 1S6, Canada. [4]Nanotechnology Research Laboratory, Faculty of Engineering Sciences, GIK Institute of Engineering Sciences and Technology, Topi 23640, Khyber Pakhtunkhwa, Pakistan.

**ABSTRACT:** In future solar cell technologies, the thermodynamic Shockley-Queisser limit for solar-to-current conversion in traditional p-n junctions could potentially be overcome with a bulk photovoltaic effect by creating an inversion broken symmetry in piezoelectric or ferroelectric materials. Here, we unveiled mechanical distortion-induced bulk photovoltaic behavior in a two-dimensional material (2D), MoTe$_2$, caused by phase transition and broken inversion symmetry in MoTe$_2$. The phase transition from single-crystalline semiconducting 2H-MoTe$_2$ to semi-metallic 1T′-MoTe$_2$ was confirmed using X-ray photoelectron spectroscopy (XPS). We used a micrometer-scale system to measure the absorption of energy, which reduced from 800 meV to 63 meV when phase transformation from hexagonal to distorted octahedral and revealed a smaller bandgap semi-metallic behavior. Experimentally, a large bulk photovoltaic response is anticipated with the maximum photovoltage $V_{OC} = 16$ mV and a positive signal of the $I_{SC} = 60$ μA (400 nm, 90.4 Wcm$^{-2}$) in the absence of an external electric field. The maximum values of both R and EQE were found to be 98 mAW$^{-1}$ and 30 %, respectively. Our findings unveil distinctive features of the photocurrent responses caused by in-plane polarity and its potential from a wide pool of established TMD-based nanomaterials, and a novel approach to reach high efficiency in converting photons-to-electricity for power harvesting optoelectronics devices.

**Corresponding author: Fahmid Kabir (Fahmid_kabir@sfu.ca)**

**INTRODUCTION**

The photovoltaic (PV) effect in traditional solar cells can be achieved by connecting two semiconductors with different doping types, which include an n-type semiconductor with an excess of electrons and a p-type semiconductor with an excess of holes. As a result, a built-in potential, which is also an intrinsic electric field, is generated at the p-n junction interface, which can be used in order to separate photo-generated, which are also light-induced, electron-hole pairs or to rectify the currents.[1, 2] There is only one component in non-centrosymmetric materials. However, a photocurrent, which is an electric current, can still exist in the absence of both built-in potential and space-time inhomogeneity.[3-6] This is referred to as the bulk PV effect.[3-5] It was discovered in the second half of the 1960s.[7] The bulk PV effect was given this name because it can be seen in non-centrosymmetric crystals under homogeneous illumination. The Shockley-Queisser (S-Q) limit, also known as the detailed balance limit, was first calculated in 1961 by William Shockley and Hans-Joachim Queisser.[6] Its maximum theoretical efficiency for a solar cell with a single p-n junction is 30% at 1.1 eV.[8] The bulk PV effect,[9] a nonlinear optical process, could give a solar cell a higher efficiency than S-Q limit.[6, 8, 10] 2D materials and non-centrosymmetric crystals lack inversion symmetry, so significant photovoltage can be generated without the use of a junction or an interface.[6, 7, 10] Achieving high efficiency with broken symmetry in 2D materials may be possible with new optoelectronic devices [4, 6, 9, 11] However, bulk PV can be generated by a single crystal under homogeneous light irradiation without the use of a p-n diode, which occurs only in materials with inversion broken symmetry or with materials that are non-centrosymmetric, which may overcome the Shockley-Queisser theoretical limit.[7, 10, 12] Bulk PV was previously discovered in ferroelectric perovskites, such as $BaTiO_3$ and $BiFeO_3$.[13, 14] However, the wideband gaps in these materials typically range between 2.7 and 5 eV, which result in insufficient absorption, low conversion efficiency, and low photocurrent density.[15-17] Innovative ferroelectric materials with narrower band gaps must be proposed in order to achieve bulk PV with a high-efficiency light-to-current conversion. Ferroelectricity in 2D materials, such as α-$In_2Se_3$ has been predicted and observed in recent years.[18, 19] Atomically thin layers of 2D materials with van der Waals (vdW) interaction have demonstrated the ability to replace cutting-edge silicon-based technology. Transition Metal Dichalcogenides (TMDs) are being studied for use in next-generation transparent conducting materials. They have higher visible & infrared light absorption, ultra-high mobility, transparency, flexibility, stability, electrical conductivity, and thermal conductivity than traditional Si and GaAs.

molybdenum ditelluride ($MoTe_2$) demonstrated room-temperature ferroelectricity,[20] and it is regarded as the most promising 2D layered nanomaterial in regards to enabling phase transition from 2H-to-1T′, which is due to a significantly small energy gap ($\Delta E < 50$ meV), among TMDs.[21, 22] Previous studies have used the shift-current model[23, 24] to explain the bulk PV effect in distorted 1T (1T‴) 2D TMDs, including $MoTe_2$.[6, 9, 10] This model predicts that the shift-current responses of bulk and monolayer 1T (1T‴) 2D TMDs exhibit strong absorption from the near-infrared (IR) to visible light region, indicating the efficient utilization of solar energy due to the intrinsic crystal anisotropy. The smaller band gaps and more delocalized valence band states in bulk 2D TMDs cause the PV effect to be significantly higher than in monolayers.[10] The bulk PV effect of 2D TMDs can be improved through the use of strain engineering, which has also been demonstrated to be a successful strategy. $MoTe_2$ with metastable phase (1T′) is a promising polar 2D material candidate without centrosymmetry that can be used for studying strong bulk PV effects according to DFT calculations, whereas hexagonal 2H-$MoTe_2$ is a semiconductor.[10, 25] Exciting physics and new booming developments beyond graphene were explored using phase transitions in 2D materials in atomically thin 2D materials from the semiconducting 2H phase, which included a trigonal prismatic structure, to the small bandgap semimetal-like 1T′ phase, which included the distorted octahedral structure.[26-28] The non-centrosymmetric space group $Pmn2_1$ of 1T′-MoTe$_2$ was discovered.[29, 30] Furthermore, the previous research indicates that the space group of the 1T'' phase is layer-thickness dependent. The even layered samples remain monoclinic, which is expected for the 1T′ phase, but their symmetry is defined by the symmorphic space group Pm.[25] In order to achieve modulated phase transitions from 2H-to-1T′ in a few layers or multi-layer MoTe$_2$, only thermal annealing,[26] which involves lasers or e-beam irradiation,[21] Chemical Vapor Deposition (CVD), such as chemical modifications, and electrostatic gating have been used up to now.[31, 32] However, these approaches frequently involve lattice distortion in treated MoTe$_2$ and material property degeneration. Using mechanical strain as a useful method in order to trigger or control the phase transition in MoTe$_2$ at room temperature in ambient environments has alternatively been adopted.[33] The metastable 1T-MoTe$_2$ and T′-MoTe$_2$ phases are difficult to synthesize, because the formation energy for the 1T′ metastable phase is greater than formation energy of the 2H phase.[32]

Strain engineering is an effective method to use in order to modify the electronic, optical, and vibrational properties of 2D materials. We investigated the phase transition of an MoTe$_2$ nanoflake from 2H to 1T′ under mechanical distortion on a PDMS substrate at 300 K in an ambient environment. The systematic Raman spectroscopy, photoelectron spectroscopy (XPS), and optical absorption spectra studies on the distorted exfoliated MoTe$_2$ thin flakes on PDMS substrate initially confirmed the phase transition induced from 2H to 1T′. We investigated the bulk PV effect, which involves a dc positive photocurrent signal being induced in a distorted MoTe$_2$ (1T′) due to the noncentrosymmetric under the illumination of the incident light, in order to demonstrate our idea. It was concluded that the photovoltage that is produced by the bulk PV device without a junction interface is slightly below the bandgap energy of MoTe$_2$ (1T′). This is on the basis of the electrical performance using the bandgap potential E$_g$/q, which q is the electric charge. Overcoming the theoretical Shockley-Queisser limit for the photon-to-current conversion in a conventional p-n junction can achieve high efficiency with broken symmetry in 2D materials for new optoelectronic devices. A distorted TMD-based nanomaterial (MoTe$_2$) exhibits bulk photovoltaic behavior. An initial structural characterization analysis could be used in order to claim that the experimental phase transition from 2H-to-1T′ was demonstrated. The energy band gap of the mechanically distorted MoTe$_2$ was measured using a micro-scale system, which was changed from 2H (800 meV) to 1T′ (63 meV). A large bulk photovoltaic response was revealed using an open-circuit voltage (V$_{OC}$ = 16 mV) and a short-circuit current (I$_{SC}$ = 60 µA) at 400 nm with an incident laser power of 90.4 Wcm$^{-2}$. The maximum value of R was discovered to be 98 mAW$^{-1}$, and the maximum EQE value was 30%. Our proposed novel strategy for inducing bulk PV in 2D materials piques the interest of researchers in solar-to-electricity conversion as a promising technology for future development.

**RESULT AND DISCUSSION**

Strain engineering can effectively tune the structural and mechanical properties of 2D materials, which notably reveals new possibilities for controlling or modifying both their electronic and optoelectronic properties.[34, 35] we used mechanical deformation to induce phase transition in MoTe$_2$ nanoflakes from 2H-MoTe$_2$ and 1T′-MoTe$_2$ to reduce the optical bandgap and break symmetry in multilayer MoTe$_2$ nanocrystals in order to induce bulk PV. Figure 1a shows a schematic of a bulk PV device configuration that is based on distorted 2H-MoTe$_2$ (1T′-MoTe$_2$) nanoflakes on a stretchable PDMS substrate, whereas Figure 1b shows an actual optical microscopy image of the device for a distorted octahedral 1T′-MoTe$_2$ on a stretched PDMS stamp.

The electrodes were deposited on the MoTe$_2$ nanoflake by ensuring that it was well aligned between the two top Au electrodes. Figure 1c depicts an AFM surface topography analysis of the MoTe$_2$ nanoflake with the corresponding step profile height showing a thickness of 10 nm, which is illustrated in Figure 1d. In regards to whether or not inhomogeneity in the distorted MoTe$_2$ is accompanied by strain variations, a net Dember effect could be observed if the formation of the photo carriers is not uniform, or an inhomogeneous photocurrent is caused by an asymmetric distribution of incident light intensity and/or sample inhomogeneity.[36] The first asymmetric light intensity profile scenario cannot exist, which is a result of the symmetry of the laser-spot profile in our setup. In order to prevent the photothermal effect, which is caused by the thermoelectric effect of laser heating, a focused laser beam with a spot size of 2 µm, a wavelength of 477 nm, and with a power limit of 100 mW/cm$^2$ was used to scan along the mirror plan between two electrodes and measure the incident-laser-position dependence of the photocurrent at different position of the flake, which is depicted in Figure 1e. The optical image of the device is displayed in the inset of Figure 1e. The sample was moved along x-axis by 2 µm (a-axis) while along y-axis it is moved by 0.5 µm (b-axis). The main response occurs when the laser spot illuminates the center of the device away from the contacts, which results in a bulk PV effect.[4] The results obtained with our devices are not suitable for the Dember effect either qualitatively or quantitatively. In addition, the photocurrent generation pattern of the device was investigated using the photocurrent mapping technique, which is shown in Figure 1f. It was demonstrated that a homogeneous photocurrent generation was seen with a small variation of 88.4 nA. However, it was discovered that the photocurrent has an antisymmetric component as a function of the incident laser beam position when a high intensity of light was used. The photothermal effect, which is caused by the thermoelectric effect of laser heating, exhibits an antisymmetric photocurrent distribution as well.[37-39] As a result, the non-BPV components in this type of case, which include the flexo-photovoltaic effect and the photothermal effect, coexist with BPV.

Figure 2 shows schematic representations of various phase structures, which include 2H-to-1T′, which are also called 1T″, transitions as well as the top and side views of various 2D phases. The structure of pristine MoTe2 is trigonal hexagonal prismatic (2H-MoTe2, which is depicted in Figure 2a, whereas the stretched MoTe2 on PDMS is distorted octahedral, which is shown in Figure 2c. The side view of the 2H phase are illustrated in Figure 2b, whereas the side view of the 1T′ phase is depicted in Figures 2d. Figure 2e depicts the initial state of a schematic band structure of 2H-MoTe2, whereas Figure 2f depicts it after the distortion. MoTe2 has an indirect band gap in its multilayered structure.[40-42] Raman spectroscopy was used in order to investigate the modulation of a 2H-to-1T′ phase transition under a tensile strain. The phase transition in the MoTe2 nanoflake from 2H to 1T′ is confirmed only after stretching, which is shown in Figure 3a. Three prominent characteristic peaks were observed for the 2H phase, which included $A_{1g}$ mode at ~174.3 cm$^{-1}$, $E_{2g}^1$ mode at ~235.9 cm$^{-1}$, and $B_{2g}^1$ mode at ~291.5 cm$^{-1}$.[40] The Raman signatures of pristine 2H-MoTe2 at ~235.9 cm$^{-1}$ and ~291.5 cm$^{-1}$ completely vanish after stretching, which is illustrated by the black curve in Figure 3a. The red curve in Figure 3a shows the Raman peaks of the 1T′ phase MoTe2 at ~119.2 cm$^{-1}$, 159.3 cm$^{-1}$, and 197.7 cm$^{-1}$. The shift in the Raman peaks was also detected during the mechanical deformation, which is shown in Supplementary Figure S1 and Supplementary Figure S2. Figure S1(e) demonstrated the observed partial reversibility of phase transition. The cause could be kinetic trap states or interface trap states between MoTe2 and PDMS as a result of the periodic wrinkling of the structure. This causes the material to exhibit partial reversibility after phase transition and causes a periodic local strain profile of alternating tensile and compressive strain even after release the PDMS. At strain levels of 1% or less, none of the phase transitions were observed. When there is a 2H strain transition of more than 1%, the mixed phase can be observed. At 1T' strain transitions of 2% or more, the mechanically induced phase transition can be seen in the material (experimental section explains how to detect strain in MoTe$_2$ crystal lattices). The distinct Raman signatures of the pristine and distorted MoTe$_2$ are attributed to the phase transition under a tensile strain. Furthermore, the observed phase transition in the TMDs materials under deformation from 2H to 1 T′ is consistent with the theoretical calculations.[22, 43] Ionic gating or chemical interactions with manifest residues in the previous studies were used in order to modulate the phase transition in 2D materials from 2H to 1T′.[44, 45] The phase change in our case was caused by the tensile strain, which instead involves a crystal structure that is free of chemical contamination and results in a clean hetero-phase transition.

We investigated the elemental composition and stoichiometry of the MoTe$_2$ films as well as changes in the crystal structure by using the X-ray Photoelectron Spectroscopy (XPS) measurement technique. The interpretation of the XPS spectra, which is shown in Figure 3b and Figure c, determines the change in the binding energy. The peaks that occur in the pristine 2H-MoTe$_2$ are observed at 228.5 eV and 573.4 eV correspond to the Mo 3d and Te 3d, whereas the peaks that occur in the distorted 1T′-MoTe$_2$ are observed at lower binding energies of 227.6 eV and 573.0 eV. It was assessed that the peaks shifted to a lower binding energy by approximately 0.9 eV and 0.4 eV, which correspond to Mo 3d and Te 3d, respectively. This can be explained by a significant shift in the Mo 3d and Te 3d peaks that are caused by different lattice symmetries in the 2H-MoTe$_2$ and 1T′-MoTe$_2$ phases, which is also relevant evidence for the phase transition from 2H-to-1T′ in the MoTe$_2$ flake under mechanical distortion. Furthermore, the absorption spectroscopy of the multi-layered MoTe$_2$ was performed using a micro-based Fournier Transform Infrared Spectroscopy (FTIR), and the optical absorption spectra of 2H-MoTe$_2$ and 1T′-MoTe$_2$ revealed absorption band edges near 800 meV and 63 meV, respectively. Our findings are consistent with a previous report, which the density functional theory simulations show that 1T′-MoTe$_2$ has a bandgap of up to 60 meV due to strong coupling between the spins and orbits in different bands.[26] The bandgap of the 2H phase is well correlated with the semiconductor behavior, whereas the bandgap of 1T′ corresponds with a small gap to the semi-metallic state of MoTe$_2$.[26] Another study discovered that 1T′-MoTe$_2$ deposited via Chemical Vapor Deposition (CVD) was compatible with its metallic nature.[46]

Figure 4a and Figure 4b show schematic representations of hexagonal 2H-MoTe$_2$ and distorted octahedral 1T′-MoTe$_2$ on PDMS substrates with Au as the source and drain electrodes. The photoconductive device, which is based on 2H-MoTe$_2$, exhibits representative current-voltage ($I_{ds}$–$V_{ds}$) curves that exhibit a photoreaction that was specified by the 45:1 photocurrent to dark current ratio at -6 mV bias, which is illustrated in Figure 4c, whereas the ($I_{ds}$–$V_{ds}$) curves on the log scale are shown in Supplementary Figure S3. Figure 4d depicts the ($I_{ds}$–$V_{ds}$) characteristics of a distorted MoTe$_2$ (2H to 1T′) device on a PDMS substrate with an effectively broken inversion symmetry.[25, 47] A typical self-biased phenomenon is observed in order to generate a current without a junction or interface on a homogeneous illumination, and the maximum photovoltage induced ($V_{oc}$ =-2.33 mV), which is obtained at $I_{ds}$ = 0 A, and a positive signal as a photocurrent response ($I_{sc}$ = 21 μA), which is obtained at $V_{ds}$ = 0 V, are analogous to the photocurrent that is induced as a bulk PV that uniquely occurs in nanocrystals with a broken inversion symmetry in the absence of external fields.[5, 48] The photocurrent is a direct current that flows through a material when it is evenly lit. It results in phenomena, such as bulk PV.[49] The bulk PV effect induced in distorted MoTe$_2$ is fairly robust both qualitatively and quantitatively. We also confirmed that the photocurrent response for the Schottky barrier photovoltaic effect at the interface of the mechanically distorted MoTe$_2$ and an Au metallic contact would not be expected. The photoconductive device based on 2H-MoTe$_2$ exhibits representative current-voltage (I$_{ds}$–V$_{ds}$) curves exhibiting photoreaction, as specified by the 45:1 photocurrent to dark current ratio at -6 mV bias (in Figure 4c), while (I$_{ds}$–V$_{ds}$) curves on the log scale are shown in Supplementary Figure S3. The mechanisms of three different light-to-current conversions are compared in Figure 4e, figure 4i, and Figure 4l, which include the photoconductive devices, photovoltaic devices, and bulk photovoltaic devices. Figure 4f depicts the typical I$_d$–V$_d$ characteristics of a photoconductive device which is not inducing of voltage by light. The current level rises on both sides under illumination, but no self-biased photovoltaic behavior is achieved, which is due to the lack of a breaking symmetry. Figure 4g and Figure 4h show a band schematic of the carrier distribution in the 2H-MoTe$_2$ channel in the dark and under illumination. The hole trap states exist in the dark at the valence band edge, which serve as a local gate. The incident photon generates photo carriers in the channel when illuminated, which results in a photocurrent and a rise in the current level of the (I$_{ds}$-V$_{ds}$) curves. Figure 4i depicts a photovoltaic device that is created by connecting two semiconductors with different doping types, which include an n-type semiconductor with an excess of electrons and a p-type semiconductor with an excess of holes. As a result, an intrinsic electric field is generated at the interface between them, which can be used in order to separate the photo-generated electron-hole pairs or to rectify the currents. The typical output characteristics show a photovoltaic effect, which is shown in Figure 4j, with a positive open-circuit voltage (V$_{oc}$) and a negative short-circuit current (I$_{sc}$) at I$_{ds}$ = 0 A. The energy band diagram that results from the Femi level alignment between two opposite semiconductors after contact, which is a result of the built-in potential at the p-n junction interface, is illustrated in Figure 4k. Figure 4l depicts the bulk PV device, which is based on the distorted octahedral MoTe$_2$, whereas Figure 4m depicts the typical output characteristics. The bulk PV can generate a photovoltage ($V_{oc}$) above the bandgap without a built-in potential, which is unlike the photovoltaic effect that is observed in solar cells.[6] This makes it prominent in regards to overcoming the Shockley-Queisser limit.[6, 50, 51] We found mechanical bulk photovoltaic behavior in single distorted $MoTe_2$ caused by phase transition, and inversion broken symmetry. Previously, the bulk PV effect was studied in a complex heterostructure where the three-fold and two-fold rotational symmetries did not match, causing the $WSe_2$ and BP heterointerfaces to lose rotational symmetry, resulting in photocurrent.[37] It has also been discovered that bulk PV effect can be found in $WS_2$-based nanotube-based devices. By gradually transitioning from a 2D monolayer ($WS_2$) to a nanotube with polar features, the bulk PV effect is greatly improved.[4] Another study used the $MoS_2/VO_2$ heterostructure to create a flexo-photovoltaic effect.[3] The charge transport analysis for the physical process is illustrated in Figure 4n by the schematic band structure of the distorted $MoTe_2$ of a simple system in a non-centrosymmetric unit cell that would exhibit the bulk PV. An arduous microscopic theory was first proposed in 1982 in order to describe the actual physics of bulk PV.[52] A photocurrent is generated in real space via electron displacements that are based on quantum transitions according to this theory. The mechanism is based on band-to-band transitions, which include inter-band excitation from the valence band to the conduction band that results in a photocurrent being caused by the charge transport or the electron displacements.[24] Photons of incident light can excite carriers across the bandgap, which allow the incident energy to be absorbed and converted into electricity via some physical interaction between the radiation and the material. The "ballistic" and "shift" mechanisms of the BPE have previously been identified from a microscopic perspective.[24] The violation of the detailed balance principle results in the ballistic mechanism, which is connected to the excitation of nonthermalized (hot) carriers in a crystal. Figure S5 shows the internal PV effect in noncentrosymmetric and centrosymmetric crystals.[24] An asymmetric momentum distribution of nonthermalized carriers in the conduction band is produced by the excitation of a photoelectron in a non-centrosymmetric crystal, which is the key distinction. The shift in space $l_0$ is caused by the nonthermalized photoexcited carriers as they lose energy and fall to the bottom of the band.[24] The BPE utilizes a quantum mechanical mechanism for its alternative shift mechanism.[13, 14] It is acquired by accounting for the nondiagonal components of the density matrix. In this instance, the virtual shift R in the real space after the carrier band-band transition is what causes the BPE rather than the carrier movement in the band. Estimated values for $l_0$ and R are in the range of 10 to 100 nm.[24]

The 1T'-MoTe$_2$ lattice has in-plane anisotropy, which is unlike most 2D materials, that have a hexagonal structure.[53] We used the edge identification method in order to determine the crystal orientations.[54] The 1T′-MoTe$_2$ has a mirror plane along the armchair direction, and the electronic polarization would appear along the direction parallel to the mirror plane, which is shown in Supplementary Figure S6. In addition, we measured the polarization angle dependence of the bulk photovoltaic effect in 1T′-MoTe$_2$ in order to determine the origin of $I_{SC}$. The schematic of the polarization-dependent photocurrent measurement set-up is shown in Figure S7a while an optical image of the device with a mirror plan exists between the $E_1$ and $E_2$ electrodes (Figure S7b). The polar diagram of a photocurrent as a function of the polarization angle is shown in Figure S7c. It showed no signs of change, and it was anisotropic along the polar direction. This result indicates that a photocurrent becomes finite even when the incident light is unpolarized, which cannot be explained solely by the BPV effect. The photovoltaic response along a specific direction in principle is allowed for linearly polarized light in noncentrosymmetric materials,[7] but a distinct polarization angle dependence, which includes sign change, is expected, and the photocurrent should vanish under unpolarized light. As a result, we can conclude that the photovoltaic effect that is caused by in-plane polarity is dominant in 1T′-MoTe$_2$. The observed photocurrent dependence on the incident laser beam polarization angle exhibits a strong in-plane anisotropy with a higher photocurrent along the mirror plane direction, and no photocurrent is observed along the direction that is perpendicular to it, which is shown in Supplementary Figure S8. The photocurrent is generated parallel to the mirror plane in our devices as opposed to being perpendicular to it, which clarified the contribution of the bulk PV effect.

Figure 5a shows the photovoltaic measurements being performed at various wavelengths of an incident laser beam, which ranged from 400 to 700 nm, using an equivalent circuit. The induced photocurrent ($I_{SC}$) vs bias voltage (V) was measured under dark and light illumination conditions. The key elements in regards to achieving a distinct self-biased phenomenon exist at a change in wavelength of the incident laser beam at a fixed input power. The bulk PV device exhibits a dominant transformation with both the open-circuit voltage ($V_{OC}$) and the dc photocurrent ($I_{SC}$) values without an external bias, which is illustrated in Figure 5b. Supplementary Figure S9 depicts the I-V curves on the log scale. More band-to-band transitions occur when illuminated with the lowest wavelength of the incident light, which activates the ($I_{SC}$ -$V_{OC}$) curves and causing the photocurrents in Figure 5b to shift to more positive values. The I-V curves demonstrate a large bulk PV response at 400 nm, which used an incident laser power of 50.3 Wcm$^{-2}$, with a $V_{OC}$ = 6.8 mV and an $I_{sc}$ = 17.5 µA, which are shown in Figure 5c. We calculated the photoresponsivity (R) and the External Quantum Efficiency (EQE) using the relationships that are provided below.[55]

$$R = \frac{I_{sc}}{P_{Laser*A}}$$

$$\text{EQE} = R\frac{hc}{e\lambda}$$

Where $I_{SC}$ is the photocurrent, $P_{Laser}$ = 50.3 W cm$^{-2}$ is the incident laser power, which $\lambda$ is its wavelength, the active area of the bulk PV device is A = 700 µm$^2$, Planck's constant is denoted by h, the speed of light is represented by c, and e represents the electron charge.[56,57] The maximum values of both R and EQE were found to be 54 mAW$^{-1}$ and 16 %. The wavelength is 400 nm, and the incident laser power is 50.3 Wcm$^{-2}$. Figure 5d shows the EQE (100%) at the fixed intensity along with different wavelengths of an incident laser beam. As the incident wavelength increases from 400 to 700 nm, R decreases sharply from 54 mAW$^{-1}$ to 10 mAW$^{-1}$. The reason is that when illuminated by incident light of a longer wavelength, the incident photons' lower energy results in a smaller number of band-to-band transitions, deactivating the ($I_{SC}$ -$V_{OC}$) curves and shifting the photocurrents to lower values (Figure 5b). In a linear plot, Figure 5b presents the dependence of $I_{SC}$ on the same laser (50.3 Wcm$^{-2}$) for seven different laser wavelengths: $I_{SC}$ (400 nm) is greater than $I_{SC}$ (650 nm), which is greater than $I_{SC}$ (500 nm). Figure S10 displays the 1T′-MoTe$_2$ absorption spectrum from 400 nm to 800 nm in wavelength range. The decrease in response with increasing wavelength could be due to variations in the wavelength-dependent absorption coefficient.[9] Also, in order to examine the photovoltaic parameters at a different input power of incident light from 12.1 to 90.4 Wcm$^{-2}$, which describes the experimental setup is displayed in the schematic of Figure 6a. The bulk PV device exhibits a dominant effect with the highest intensity of incident light. It activates the ($I_{SC}$-$V_{OC}$) curves to move to more positive values of the photocurrents, which are displayed in Figure 6b. The I -V curves on the log scale can be seen in Supplementary Figure S11. The large bulk PV caused intrinsic transitions, which were band to band, at various input intensities when the laser beam (400 nm) was incident, and both $I_{SC}$ and $V_{OC}$ increased monotonically as the power of the incident laser beam was increased. The I-V curves expose a large bulk PV response at an incident laser power of 90.4 W cm$^{-2}$ (400 nm) with $V_{oc}$ = 16 mV and photocurrent $I_{sc}$ = 60 μA, which is shown in Figure 6c. The maximum values for both R and EQE were found to be 98 mAW$^{-1}$ and 30 %, which included a (wavelength of 400 nm and an incident laser power of 90.4 Wcm$^{-2}$. Figure 6d shows the EQE (100%) as a function of the wavelength at a fixed intensity of different lights. The power conversion efficiency (η) of bulk PV device is calculated at power of 90.4 Wcm$^{-2}$ by using the following relations.[58]

$$\text{Fill Factor=FF} = I_m V_m / I_{sc} V_{oc}$$

And

$$PCE = \eta = V_{oc} J_{sc}/P_{laser} \times FF \times 100$$

The parameters ($V_{oc} = 16\ mV$), ($I_{sc} = 60\ \mu A$), ($V_m = 9.16\ mV$), and ($I_m = 9.16\ \mu A$) are estimated from Figure S12. The illuminated device's area is 700 μm$^2$. The calculated value of FF is 0.30. The device's PCE is calculated to be 0.40%. We calculated the specific detectivity ($D$) of the device by using the relation as follow:[59]

$$D = R \Big/ \sqrt{2J_d}$$

where R and $J_d$ are responsivity and dark current density, respectively. Our device displayed a specific detectivity value for $2.2 \times 10^{-1}$ jones. Figure S13 depicts the measurement of time-resolved response using a laser with a power of 50.3 Wcm$^{-2}$ and a wavelength of 400 nm. The on-off ratio of as a photodetector was calculated to be 20.6. Given that all reported papers on 2D TMDs, including ours for examining bulk PV, focus on fundamental studies.[6] Because we used nanoflakes exfoliated from natural bulk crystals, our photoconversion efficiency is less than 1%. Our research on the bulk PV effect in distorted MoTe$_2$ nanoflakes, on the other hand, may provide a method for attempting to implement it for large area 2D TMD thin films in the future.[6] Currently, however, efforts are focused on creating large-area 2D materials with a smooth morphology, high transparency, and sufficient energy levels. The high cost of 2D materials is a significant barrier to their long-term commercialization. High-performance PV technologies are anticipated to be developed at an industrial level with the help of creative, affordable design techniques for 2D materials. In the revised manuscript, we revised our statement about the S-Q limit. Our findings suggest that it could be possible to propose optoelectronic devices in the future that are tuned via mechanical distortion in 2D materials, which would enable a novel approach in regards to simulating interactions between matter and light.

**CONCLUSIONS**

In conclusion, overcoming the theoretical S-Q limit for photon-to-current conversion in a conventional p-n junction may enable new optoelectronic devices to achieve high efficiency in 2D materials with broken symmetry. In this paper, we present a novel charge separation and photovoltage generation mechanism that only operates at the nanoscale and in the presence of visible light. We investigated the bulk photovoltaic behavior of a distorted 2D TMD-based nanomaterial ($MoTe_2$). The experiential phase transition from 2H-to-1T′ could be claimed to have been demonstrated using an initial structural characterization analysis using XPS and Raman spectroscopy. We implemented a micro-scale-based system to measure the energy band gap of mechanically distorted $MoTe_2$, changed from 2H (800 meV) to 1T′ with an energy band gap of 63 meV. The photocurrent mapping technique was used to examine the device's photocurrent generation pattern. The present results propose a simple approach to creating new optoelectronic devices with strain gradient phenomena. Based on our findings, ultrathin 2D TMDs may be capable of producing high-efficiency solar cells, but further development is required.

**METHODS**

Natural bulk high-quality 2D crystals of MoTe$_2$, which include a (2H phase-semiconductor, with a typical lateral size of ~0.6-0.8 cm are commercially available at HQ graphene. Layered MoTe$_2$ material was broken up into atomically thin few-layer nanoflakes using a top-down method that is called dry mechanical exfoliation on a stretchable PDMS substrate in a controlled environment. The strain-modulation of the phase transition of MoTe$_2$ is found to be significantly less tensile strain at 2% while the PDMS is stressed up to 10%. The experimental setup that was used to stretch the MoTe$_2$ device is shown in Figure S14. Strain, which is defined as the total deformation (elongation) per reference length of a material as a result of some applied stress, is a dimensionless quantity that expresses a proportional dimensional change and calculated by using the relation $\varepsilon = \Delta\ell/\ell_0$, (for pristine MoTe$_2$ $\ell_0$= 20 µm), (for distorted MoTe$_2$ $\Delta\ell$ = 0.4 µm). Dark-field microscopy was used to monitor the MoTe$_2$ nanoflakes in order to achieve a uniform layer thickness, and it measures the elongation of distorted MoTe$_2$. Raman Spectroscopy (RS), XPS spectra, and absorption spectroscopy techniques were initially well-suited for the structural characterization of MoTe$_2$ before (2H-MoTe$_2$) and after distorted (1T′-MoTe$_2$). We used a commercial setup with an excitation laser source, which included hv = 2.41 eV, λ= 514 nm, and 1800/mm grating, with a working distance of (100×) to excite the sample while collecting the scattered photons in order to confirm the crystal structures and detect the Raman signatures of MoTe$_2$ (i.e., 2H-to-1T′). The incident laser beam's power was kept constant at 50 mW with a spot size of 0.7 µm. A micro-focused monochromatic setup (Thermo Scientific K-AlphaTM+) was used in order to characterize the binding energies of 2H-to-1T′ MoTe$_2$ for the X-ray Photoelectron Spectroscopy (XPS). The electrodes were finally transferred onto the onto MoTe$_2$ flake for the electrical transport measurements. The electrical characterizations were conducted in a vacuum box at room temperature (300K) using a Keithley series 2400 source meter and a 5½-digit model 6485 Pico ammeter. The bulk PV device was illuminated at a fixed intensity using a light source with a broadband spectral range with a wavelength from 400 to 700 nm and an incident laser power of 50.3 W cm$^{-2}$ for the photovoltaic parameters. The same bulk PV device was also illuminated using varying incident light intensities with an incident laser power that ranged from 12.1 to 90.4 W cm$^{-2}$ at a fixed wavelength of 400 nm. A focused 477 nm laser beam was used to raster-scan the sample, and a Keithley 2400 Source Meter was used to scan the photocurrent measurements at various points in order to map the photocurrent generation of the distorted MoTe$_2$ device. The laser beam's power was limited in order to not exceed the solar intensity of 100 mW/cm$^2$.

**ASSOCIATED CONTENT**

**Supporting Information**

The Supporting Information is available free of charge at

Schematic illustrations and Raman spectra of 2H-MoTe$_2$, and 1T′-MoTe$_2$; hexagonal lattice structures of 2H-MoTe$_2$ and 1T′-MoTe$_2$; $I_{ds}$–$V_{ds}$ characteristics; $I_{ds}$–$V_{ds}$ characteristics of different BPVE devices; non-centrosymmetric and centrosymmetric crystals representing the bulk and classical photovoltaic effects; atomic arrangement of a distorted T′-MoTe$_2$ in top view to locate the mirror plan; schematic of the polarization-dependent photocurrent measurement set-up; photocurrent's dependence on the incident laser beam's polarization angle; $I_{ds}$–$V_{ds}$ characteristics on log scale; absorption spectrum of 1T′-MoTe$_2$; time-resolved response; experimental setup to stretch the MoTe$_2$ device

**ACKNOWLEDGMENTS**

This work was supported by the National Research Foundation of Korea (NRF) grant funded by the Korea government (MSIT) (No. RS-2022-00165798).

**Competing interests**

The authors declare no competing interests.

**REFERENCES**
(1) Frisenda, R.; Molina-Mendoza, A. J.; Mueller, T.; Castellanos-Gomez, A.; van der Zant, H. S. J. Atomically thin p–n junctions based on two-dimensional materials. *Chemical Society Reviews* **2018**, *47* (9), 3339-3358, 10.1039/C7CS00880E.
(2) Aftab, S.; Akhtar, I.; Seo, Y.; Eom, J. WSe2 homojunction p–n diode formed by photoinduced activation of mid-gap defect states in boron nitride. *ACS applied materials & interfaces* **2020**, *12* (37), 42007-42015.
(3) Jiang, J.; Chen, Z.; Hu, Y.; Xiang, Y.; Zhang, L.; Wang, Y.; Wang, G.-C.; Shi, J. Flexo-photovoltaic effect in MoS2. *Nature Nanotechnology* **2021**, *16* (8), 894-901.
(4) Zhang, Y.; Ideue, T.; Onga, M.; Qin, F.; Suzuki, R.; Zak, A.; Tenne, R.; Smet, J.; Iwasa, Y. Enhanced intrinsic photovoltaic effect in tungsten disulfide nanotubes. *Nature* **2019**, *570* (7761), 349-353.
(5) Nadupalli, S.; Kreisel, J.; Granzow, T. Increasing bulk photovoltaic current by strain tuning. *Science advances* **2019**, *5* (3), eaau9199.
(6) Aftab, S.; Iqbal, M. Z.; Haider, Z.; Iqbal, M. W.; Nazir, G.; Shehzad, M. A. Bulk Photovoltaic Effect in 2D Materials for Solar-Power Harvesting. *Advanced Optical Materials* **2022**, 2201288.
(7) Sturman, B. I.; Fridkin, V. M. *The photovoltaic and photorefractive effects in noncentrosymmetric materials*; Routledge, 2021.
(8) Shockley, W.; Queisser, H. J. Detailed Balance Limit of Efficiency of p-n Junction Solar Cells. *Journal of Applied Physics* **1961**, *32* (3), 510-519.
(9) Yadgarov, L.; Višić, B.; Abir, T.; Tenne, R.; Polyakov, A. Y.; Levi, R.; Dolgova, T. V.; Zubyuk, V. V.; Fedyanin, A. A.; Goodilin, E. A. Strong light–matter interaction in tungsten disulfide nanotubes. *Physical Chemistry Chemical Physics* **2018**, *20* (32), 20812-20820.
(10) Ai, H.; Kong, Y.; Liu, D.; Li, F.; Geng, J.; Wang, S.; Lo, K. H.; Pan, H. 1T‴ Transition-Metal Dichalcogenides: Strong Bulk Photovoltaic Effect for Enhanced Solar-Power Harvesting. *The Journal of Physical Chemistry C* **2020**, *124* (20), 11221-11228.
(11) The Quarter-Century Anniversary of Carbon Nanotube Research. *ACS nano* **2017**, *11* (1), 1-2.
(12) von Baltz, R.; Kraut, W. Theory of the bulk photovoltaic effect in pure crystals. *Physical Review B* **1981**, *23* (10), 5590.
(13) Young, S. M.; Rappe, A. M. First principles calculation of the shift current photovoltaic effect in ferroelectrics. *Physical review letters* **2012**, *109* (11), 116601.
(14) Young, S. M.; Zheng, F.; Rappe, A. M. First-principles calculation of the bulk photovoltaic effect in bismuth ferrite. *Physical review letters* **2012**, *109* (23), 236601.
(15) Grinberg, I.; West, D. V.; Torres, M.; Gou, G.; Stein, D. M.; Wu, L.; Chen, G.; Gallo, E. M.; Akbashev, A. R.; Davies, P. K. Perovskite oxides for visible-light-absorbing ferroelectric and photovoltaic materials. *Nature* **2013**, *503* (7477), 509-512.
(16) Han, H.; Song, S.; Lee, J. H.; Kim, K. J.; Kim, G.-W.; Park, T.; Jang, H. M. Switchable photovoltaic effects in hexagonal manganite thin films having narrow band gaps. *Chemistry of Materials* **2015**, *27* (21), 7425-7432.

(17) Bai, Y.; Jantunen, H.; Juuti, J. Ferroelectric Oxides for Solar Energy Conversion, Multi-Source Energy Harvesting/Sensing, and Opto-Ferroelectric Applications. *ChemSusChem* **2019**, *12* (12), 2540-2549.
(18) Shirodkar, S. N.; Waghmare, U. V. Emergence of ferroelectricity at a metal-semiconductor transition in a 1 T monolayer of MoS 2. *Physical review letters* **2014**, *112* (15), 157601.
(19) Ding, W.; Zhu, J.; Wang, Z.; Gao, Y.; Xiao, D.; Gu, Y.; Zhang, Z.; Zhu, W. Prediction of intrinsic two-dimensional ferroelectrics in In2Se3 and other III2-VI3 van der Waals materials. *Nature communications* **2017**, *8* (1), 1-8.
(20) Yuan, S.; Luo, X.; Chan, H. L.; Xiao, C.; Dai, Y.; Xie, M.; Hao, J. Room-temperature ferroelectricity in MoTe2 down to the atomic monolayer limit. *Nature communications* **2019**, *10* (1), 1-6.
(21) Cho, S.; Kim, S.; Kim, J. H.; Zhao, J.; Seok, J.; Keum, D. H.; Baik, J.; Choe, D.-H.; Chang, K. J.; Suenaga, K. Phase patterning for ohmic homojunction contact in MoTe2. *Science* **2015**, *349* (6248), 625-628.
(22) Duerloo, K.-A. N.; Li, Y.; Reed, E. J. Structural phase transitions in two-dimensional Mo-and W-dichalcogenide monolayers. *Nature communications* **2014**, *5* (1), 1-9.
(23) Pal, S.; Sarath, N.; Priya, K. S.; Murugavel, P. A review on ferroelectric systems for next generation photovoltaic applications. *Journal of Physics D: Applied Physics* **2022**, *55* (28), 283001.
(24) Zenkevich, A.; Matveyev, Y.; Maksimova, K.; Gaynutdinov, R.; Tolstikhina, A.; Fridkin, V. Giant bulk photovoltaic effect in thin ferroelectric BaTiO 3 films. *Physical Review B* **2014**, *90* (16), 161409.
(25) Beams, R.; Cançado, L. G.; Krylyuk, S.; Kalish, I.; Kalanyan, B.; Singh, A. K.; Choudhary, K.; Bruma, A.; Vora, P. M.; Tavazza, F. Characterization of few-layer 1T′ MoTe2 by polarization-resolved second harmonic generation and Raman scattering. *ACS nano* **2016**, *10* (10), 9626-9636.
(26) Keum, D. H.; Cho, S.; Kim, J. H.; Choe, D.-H.; Sung, H.-J.; Kan, M.; Kang, H.; Hwang, J.-Y.; Kim, S. W.; Yang, H.; et al. Bandgap opening in few-layered monoclinic MoTe2. *Nature Physics* **2015**, *11* (6), 482-486.
(27) Tan, Y.; Luo, F.; Zhu, M.; Xu, X.; Ye, Y.; Li, B.; Wang, G.; Luo, W.; Zheng, X.; Wu, N. Controllable 2H-to-1T′ phase transition in few-layer MoTe 2. *Nanoscale* **2018**, *10* (42), 19964-19971.
(28) Hughes, H.; Friend, R. Electrical resistivity anomaly in β-MoTe2 (metallic behaviour). *Journal of Physics C: Solid State Physics* **1978**, *11* (3), L103.
(29) Zhang, K.; Bao, C.; Gu, Q.; Ren, X.; Zhang, H.; Deng, K.; Wu, Y.; Li, Y.; Feng, J.; Zhou, S. Raman signatures of inversion symmetry breaking and structural phase transition in type-II Weyl semimetal MoTe2. *Nature Communications* **2016**, *7* (1), 13552.
(30) Tamai, A.; Wu, Q.; Cucchi, I.; Bruno, F. Y.; Riccò, S.; Kim, T. K.; Hoesch, M.; Barreteau, C.; Giannini, E.; Besnard, C. Fermi arcs and their topological character in the candidate type-II Weyl semimetal MoTe 2. *Physical Review X* **2016**, *6* (3), 031021.
(31) Empante, T. A.; Zhou, Y.; Klee, V.; Nguyen, A. E.; Lu, I.-H.; Valentin, M. D.; Naghibi Alvillar, S. A.; Preciado, E.; Berges, A. J.; Merida, C. S. Chemical vapor deposition growth of few-layer MoTe2 in the 2H, 1T′, and 1T phases: tunable properties of MoTe2 films. *ACS nano* **2017**, *11* (1), 900-905.
(32) Sokolikova, M. S.; Mattevi, C. Direct synthesis of metastable phases of 2D transition metal dichalcogenides. *Chemical Society Reviews* **2020**, *49* (12), 3952-3980.
(33) Song, S.; Keum, D. H.; Cho, S.; Perello, D.; Kim, Y.; Lee, Y. H. Room Temperature Semiconductor–Metal Transition of MoTe2 Thin Films Engineered by Strain. *Nano letters* **2016**, *16* (1), 188-193.
(34) Deng, S.; Sumant, A. V.; Berry, V. Strain engineering in two-dimensional nanomaterials beyond graphene. *Nano Today* **2018**, *22*, 14-35.
(35) Johari, P.; Shenoy, V. B. Tuning the electronic properties of semiconducting transition metal dichalcogenides by applying mechanical strains. *ACS nano* **2012**, *6* (6), 5449-5456.
(36) Tauc, J. Generation of an emf in semiconductors with nonequilibrium current carrier concentrations. *Reviews of Modern Physics* **1957**, *29* (3), 308.

(37) Akamatsu, T.; Ideue, T.; Zhou, L.; Dong, Y.; Kitamura, S.; Yoshii, M.; Yang, D.; Onga, M.; Nakagawa, Y.; Watanabe, K. A van der Waals interface that creates in-plane polarization and a spontaneous photovoltaic effect. *Science* **2021**, *372* (6537), 68-72.
(38) Osterhoudt, G. B.; Diebel, L. K.; Gray, M. J.; Yang, X.; Stanco, J.; Huang, X.; Shen, B.; Ni, N.; Moll, P. J. W.; Ran, Y.; et al. Colossal mid-infrared bulk photovoltaic effect in a type-I Weyl semimetal. *Nature materials* **2019**, *18* (5), 471-475.
(39) Brody, P. S. High voltage photovoltaic effect in barium titanate and lead titanate-lead zirconate ceramics. *Journal of Solid State Chemistry* **1975**, *12* (3), 193-200.
(40) Aftab, S.; Khan, M. F.; Gautam, P.; Noh, H.; Eom, J. MoTe 2 van der Waals homojunction p–n diode with low resistance metal contacts. *Nanoscale* **2019**, *11* (19), 9518-9525.
(41) Lezama, I. G.; Arora, A.; Ubaldini, A.; Barreteau, C.; Giannini, E.; Potemski, M.; Morpurgo, A. F. Indirect-to-direct band gap crossover in few-layer MoTe2. *Nano letters* **2015**, *15* (4), 2336-2342.
(42) Lezama, I. G.; Ubaldini, A.; Longobardi, M.; Giannini, E.; Renner, C.; Kuzmenko, A. B.; Morpurgo, A. F. Surface transport and band gap structure of exfoliated 2H-MoTe2 crystals. *2D Materials* **2014**, *1* (2), 021002.
(43) Yamamoto, M.; Wang, S. T.; Ni, M.; Lin, Y.-F.; Li, S.-L.; Aikawa, S.; Jian, W.-B.; Ueno, K.; Wakabayashi, K.; Tsukagoshi, K. Strong enhancement of Raman scattering from a bulk-inactive vibrational mode in few-layer MoTe2. *ACS nano* **2014**, *8* (4), 3895-3903.
(44) Kappera, R.; Voiry, D.; Yalcin, S. E.; Branch, B.; Gupta, G.; Mohite, A. D.; Chhowalla, M. Phase-engineered low-resistance contacts for ultrathin MoS2 transistors. *Nature materials* **2014**, *13* (12), 1128-1134.
(45) Wang, Y.; Xiao, J.; Zhu, H.; Li, Y.; Alsaid, Y.; Fong, K. Y.; Zhou, Y.; Wang, S.; Shi, W.; Wang, Y. Structural phase transition in monolayer MoTe2 driven by electrostatic doping. *Nature* **2017**, *550* (7677), 487-491.
(46) Pace, S.; Martini, L.; Convertino, D.; Keum, D. H.; Forti, S.; Pezzini, S.; Fabbri, F.; Mišeikis, V.; Coletti, C. Synthesis of Large-Scale Monolayer 1T′-MoTe2 and Its Stabilization via Scalable hBN Encapsulation. *ACS nano* **2021**, *15* (3), 4213-4225.
(47) Chen, S.-Y.; Goldstein, T.; Venkataraman, D.; Ramasubramaniam, A.; Yan, J. Activation of new Raman modes by inversion symmetry breaking in type II Weyl semimetal candidate T′-MoTe2. *Nano letters* **2016**, *16* (9), 5852-5860.
(48) Yang, M.-M.; Kim, D. J.; Alexe, M. Flexo-photovoltaic effect. *Science* **2018**, *360* (6391), 904-907.
(49) Rangel, T.; Fregoso, B. M.; Mendoza, B. S.; Morimoto, T.; Moore, J. E.; Neaton, J. B. Large bulk photovoltaic effect and spontaneous polarization of single-layer monochalcogenides. *Physical review letters* **2017**, *119* (6), 067402.
(50) Morimoto, T.; Nagaosa, N. Topological aspects of nonlinear excitonic processes in noncentrosymmetric crystals. *Physical Review B* **2016**, *94* (3), 035117.
(51) Spanier, J. E.; Fridkin, V. M.; Rappe, A. M.; Akbashev, A. R.; Polemi, A.; Qi, Y.; Gu, Z.; Young, S. M.; Hawley, C. J.; Imbrenda, D. Power conversion efficiency exceeding the Shockley–Queisser limit in a ferroelectric insulator. *Nature Photonics* **2016**, *10* (9), 611-616.
(52) Belinicher, V.; Ivchenko, E.; Sturman, B. Kinetic theory of the displacement photovoltaic effect in piezoelectrics. *Zh. Eksp. Teor. Fiz* **1982**, *83* (2), 649.
(53) Song, Q.; Wang, H.; Pan, X.; Xu, X.; Wang, Y.; Li, Y.; Song, F.; Wan, X.; Ye, Y.; Dai, L. Anomalous in-plane anisotropic Raman response of monoclinic semimetal 1T´-MoTe2. *Scientific reports* **2017**, *7* (1), 1758.
(54) Guo, Y.; Liu, C.; Yin, Q.; Wei, C.; Lin, S.; Hoffman, T. B.; Zhao, Y.; Edgar, J. H.; Chen, Q.; Lau, S. P.; et al. Distinctive in-Plane Cleavage Behaviors of Two-Dimensional Layered Materials. *ACS nano* **2016**, *10* (9), 8980-8988.

(55) Aftab, S.; Samiya, M.; Iqbal, M. W.; Shinde, P.; Khan, M. F.; ur Rehman, A.; Yousaf, S.; Park, S.; Jun, S. C. Two-Dimensional Electronic Devices Modulated by the Activation of Donor-Like States in Boron Nitride. *Nanoscale* **2020**.
(56) Furchi, M. M.; Pospischil, A.; Libisch, F.; Burgdörfer, J.; Mueller, T. Photovoltaic effect in an electrically tunable van der Waals heterojunction. *Nano letters* **2014**, *14* (8), 4785-4791.
(57) Ye, L.; Li, H.; Chen, Z.; Xu, J. Near-infrared photodetector based on MoS2/black phosphorus heterojunction. *Acs Photonics* **2016**, *3* (4), 692-699.
(58) Aftab, S.; Iqbal, M. Z.; Alam, S.; Alzaid, M. Effect of an optimal oxide layer on the efficiency of graphene-silicon Schottky junction solar cell. *International Journal of Energy Research* **2021**, *45* (12), 18173-18181.
(59) Aftab, S.; Samiya, M.; Haq, H. M. U.; Iqbal, M. W.; Hussain, M.; Yousuf, S.; Rehman, A. U.; Khan, M. U.; Ahmed, Z.; Iqbal, M. Z. Single nanoflake-based PtSe 2 p–n junction (in-plane) formed by optical excitation of point defects in BN for ultrafast switching photodiodes. *Journal of Materials Chemistry C* **2021**, *9* (1), 199-207.

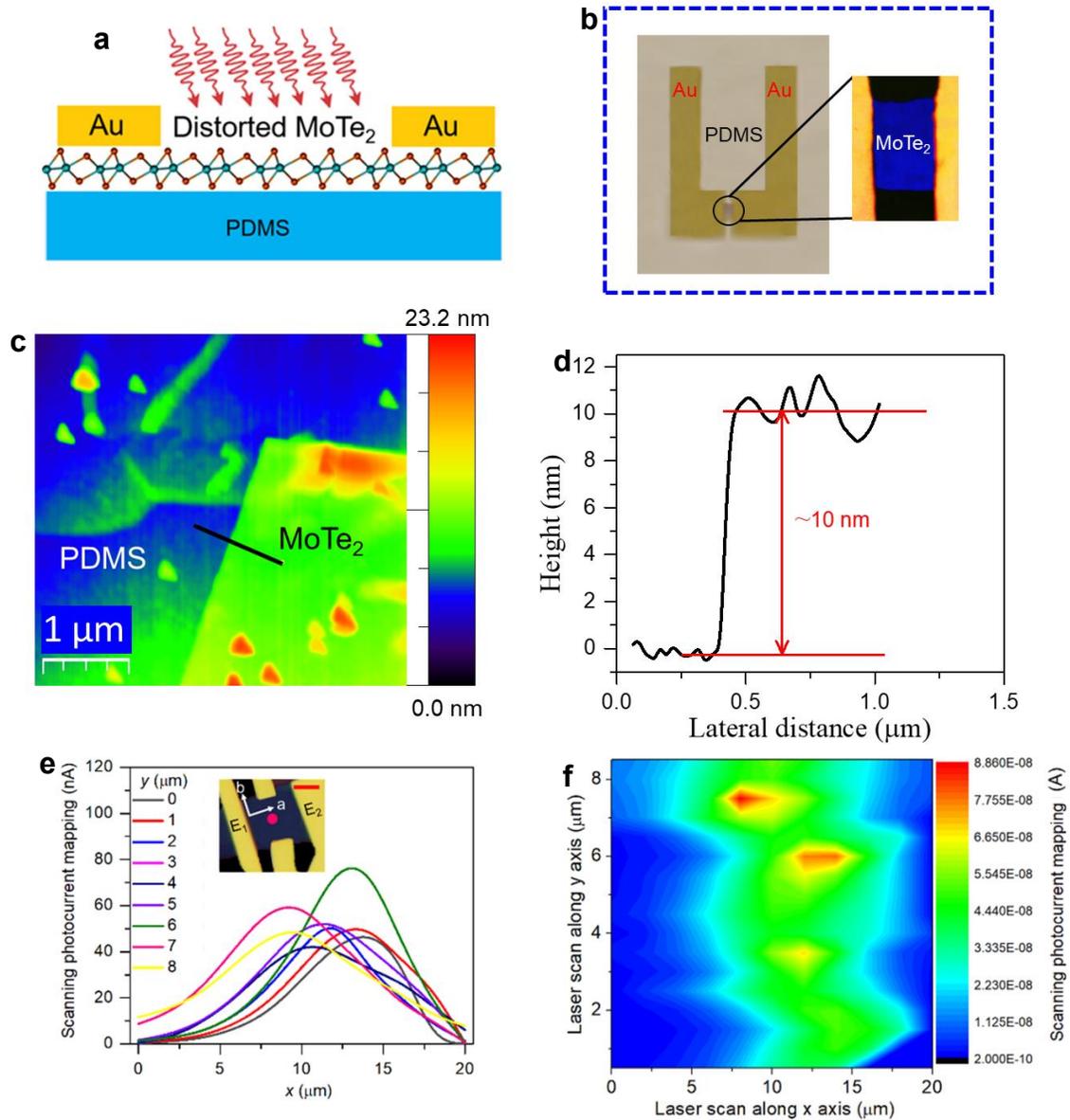

**Figure 1. The schematic and AFM topography of distorted MoTe₂.** (a) A schematic illustration of a distorted 2H-MoTe₂ (1T′-MoTe₂) nanoflake on the stretchable PDMS substrate. (b) An optical microscope image of the sample on the stretched PDMS substrate with Au electrodes deposited as electrical contact pads. The inset of b is an enhanced dark field optical microscopy image of a distorted nanoflake on PDMS. The distance between the electrodes is 20 µm. (c) AFM topography of the active area of the device. (d)The profile height shows the thickness of the nanoflake ~10 nm, which was scanned from the AFM image in Figure c. (e) The photocurrent measured as a function of the position of the laser beam at different position of the flake. The sample was moved along x-axis by 2 µm (a-axis) while along y-axis it is moved by 0.5 µm (b-axis). A focused 477 nm laser beam was used to raster-scan over the sample, whereas a Keithley 2400 Source Meter was used to scan photocurrent measurements at different points in order to map the photocurrent

generation of the distorted MoTe$_2$ device. The laser beam's power was limited in order to not exceed the solar intensity (100 mW /cm$^2$). The inset of e shows the optical image of the device. The red scale bar is 10 µm.   (f)  The pattern of scanning photocurrent generation.

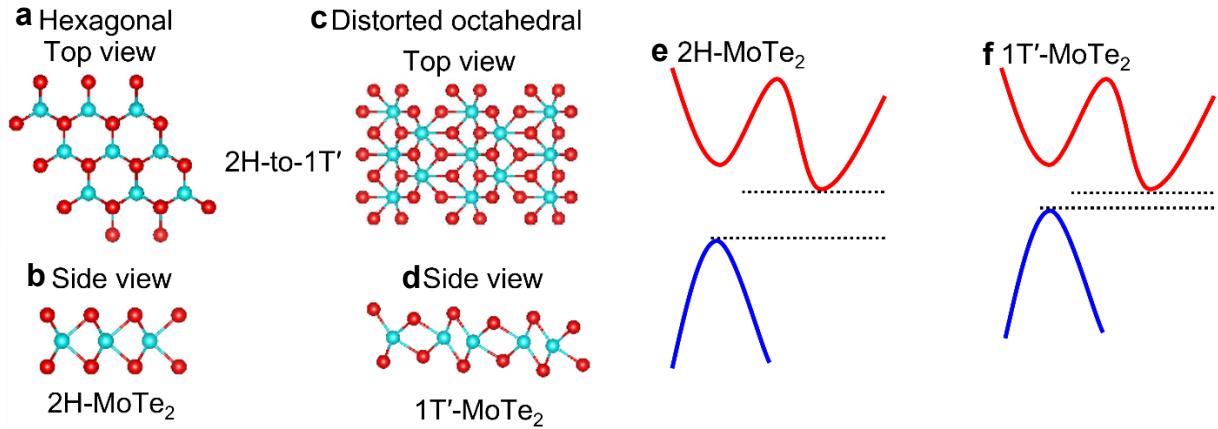

**Figure 2. Electronic band structures of MoTe$_2$.** (a) Top view of the hexagonal lattice structures of 2H-MoTe$_2$ in its initial state. (b) The side view of the same structure. The blue spheres represent the Mo atoms, and the red spheres represent the Te atoms. (c) The top view of the octahedral lattice structures of 1T′-MoTe$_2$. (d) The side view of the same structure. (e) Schematic band structure of pristine MoTe$_2$. (f) Schematic band structure of distorted MoTe$_2$.

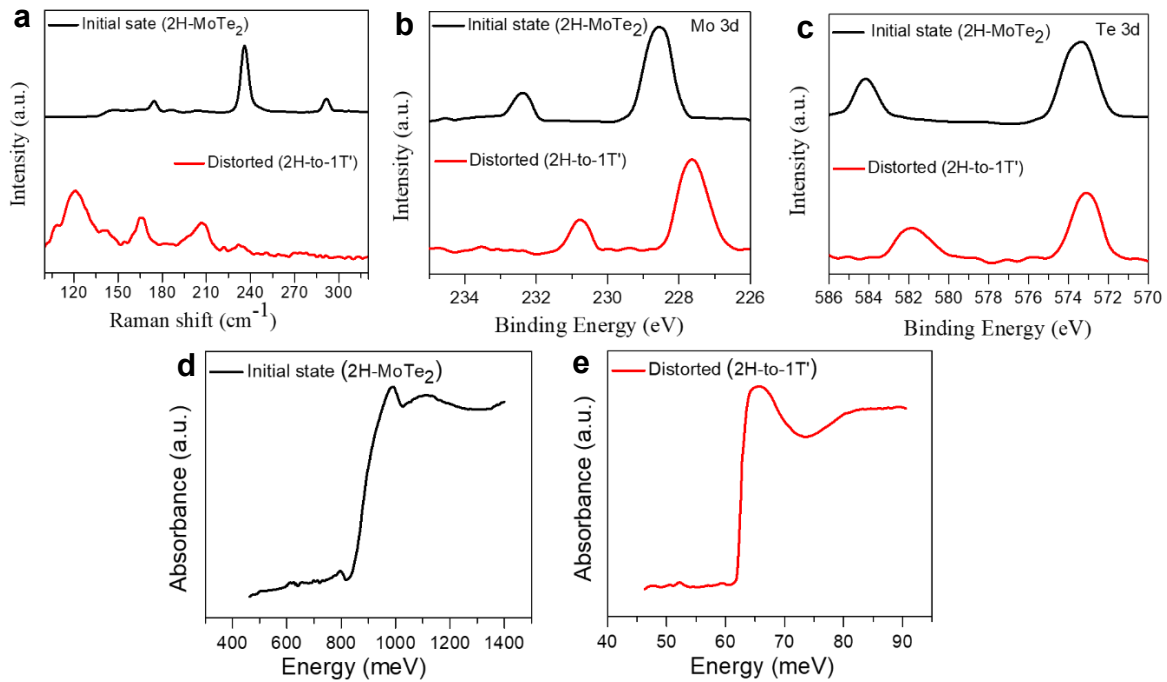

**Figure 3 Characterization of phase transition in MoTe$_2$.** (a) UV-Raman spectroscopy of 2H, which is illustrated by the black curve, and the metastable 1T′ phases, which is illustrated by the red curve, of the same MoTe$_2$ nanoflake before and after the distorted display phase transitions from 2H-to-1T′. A typical XPS spectra of 2H and 1T′-MoTe$_2$ nanoflakes. (b) The Mo 3d initial state, which is illustrated by the black curve. (c) Te 3d after distortion, which is illustrated by the red curve. (d) 2H-MoTe$_2$ and **e** 1T′-MoTe$_2$ micro-based FTIR absorption spectroscopy.

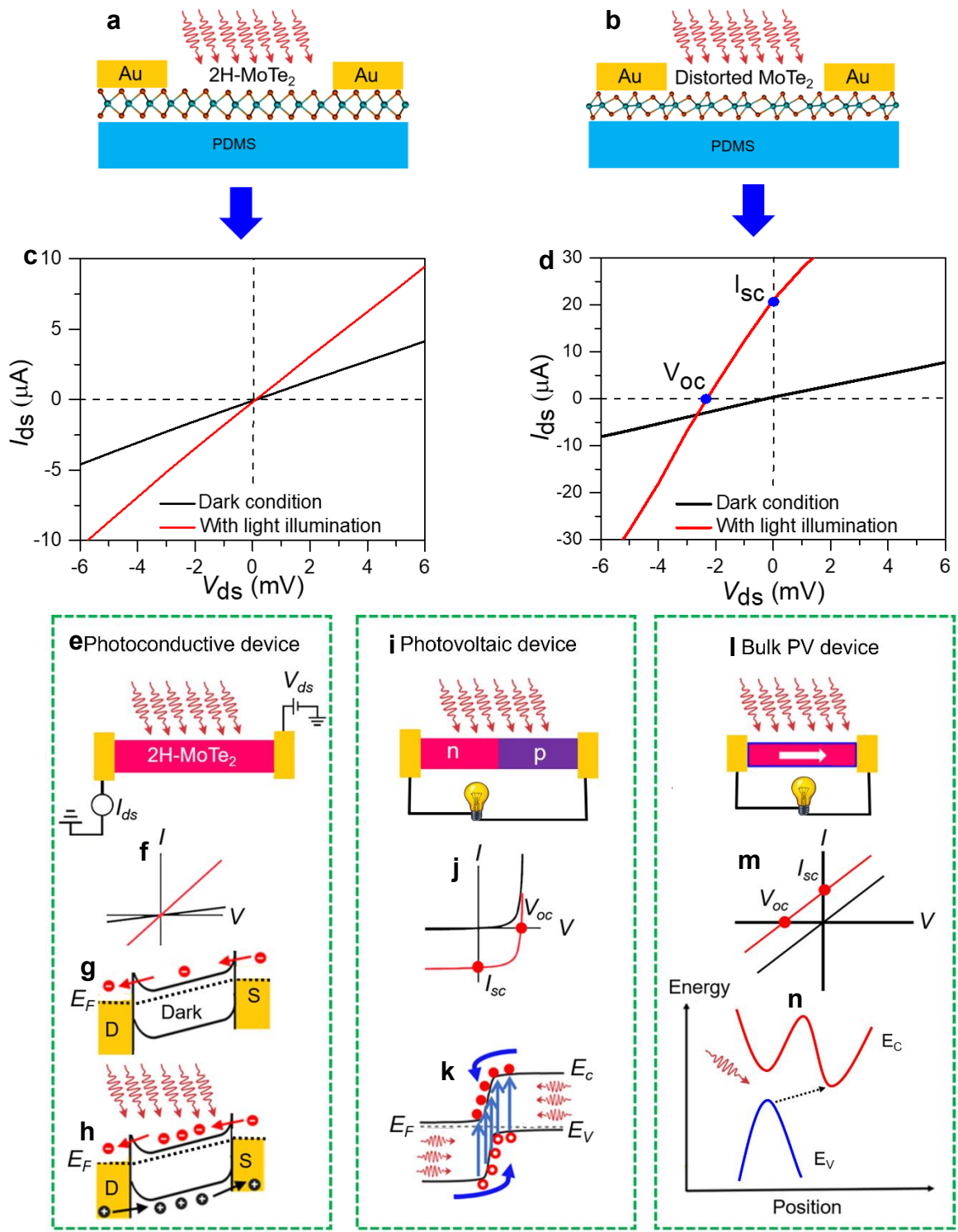

**Figure 4. Electrical properties of MoTe$_2$.** Schematic illustrations of 2H-MoTe$_2$ on the PDMS substrate under the illumination of light with Pd/Au electrodes (a) before and (b) after stretching. I$_{ds}$–V$_{ds}$ characteristics of (c) of 2H-MoTe$_2$ nanoflake demonstrating photoconductive effect. (d) distorted 2H-MoTe$_2$ (1T′-MoTe$_2$) demonstrating induced photocurrent under visible light illumination (50.3 Wcm$^{-2}$, 600 nm). Comparison of three different light-to-current conversion

mechanisms: (e) Photoconductive device schematic; (f) typical I–V characteristics without self-biased photovoltaic behavior; g band alignment for a 2H-MoTe$_2$ channel contacted with two metal electrodes in the dark under external bias; and h band alignment under illumination. (i) Photovoltaic device schematic; (j) typical I–V characteristics demonstrating the mechanism of a photovoltaic effect; and k energy band diagram after contact with two opposite semiconductors demonstrating the alignment of Femi levels as a result of the formation of the p-n junction. (l) A schematic illustration of a bulk PV device; (m) typical I–V characteristics demonstrating a bulk PV effect mechanism; and n a schematic band structure of distorted MoTe$_2$ that would exhibit the bulk PV effect.

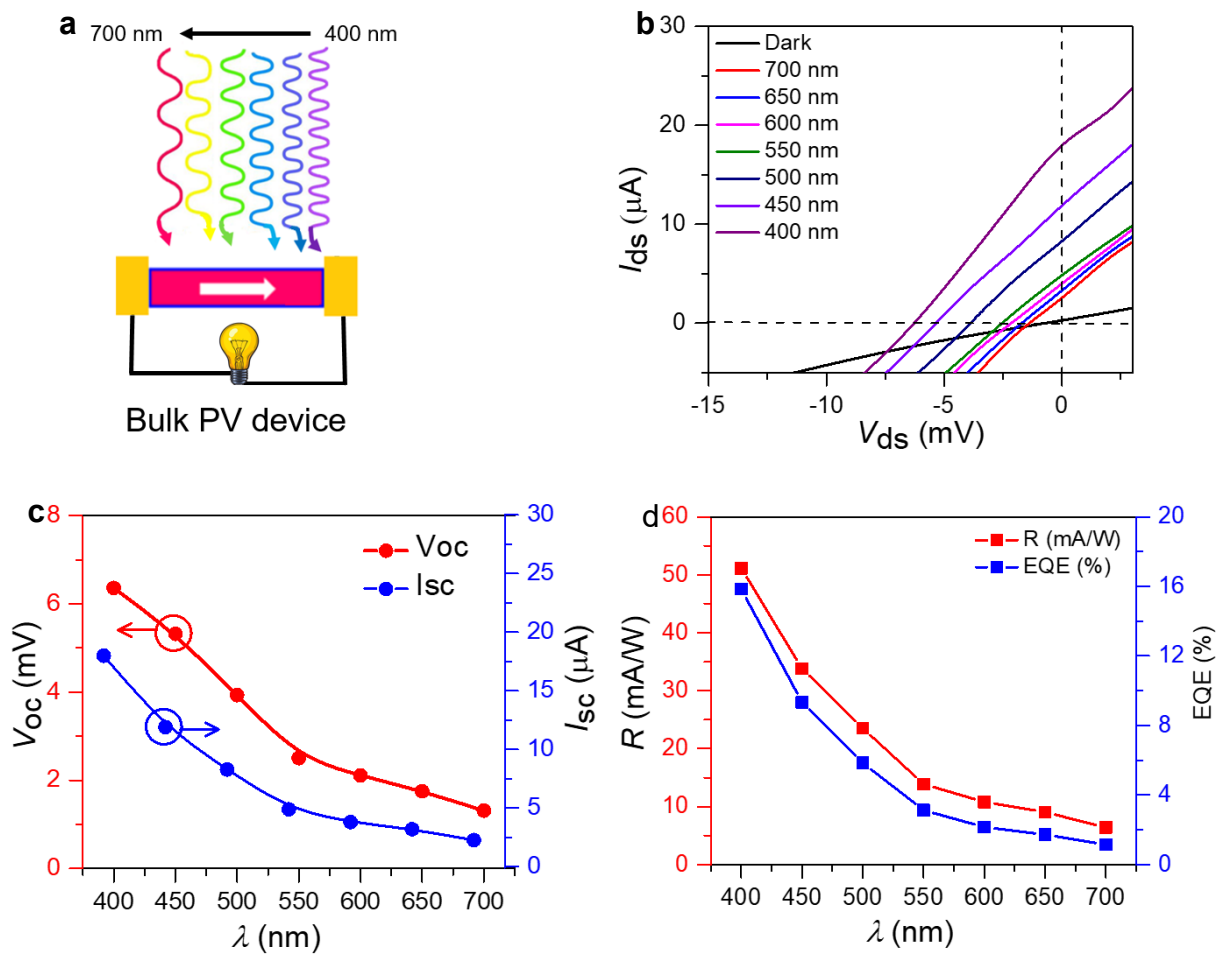

**Figure 5. Electrical characteristics of distorted MoTe$_2$ under incident light of various wavelengths.** (a) Schematic illustrations show bulk PV devices under illumination at different wavelengths. (b) $I_{ds}$–$V_{ds}$ characteristics were recorded at a different wavelength of laser illumination at a fixed intensity of incident light (50.3 Wcm$^{-2}$). (c) Dependence of both Voc and Isc on different wavelengths of laser illumination. (d) Dependence of R and EQE on the wavelength of incident light.

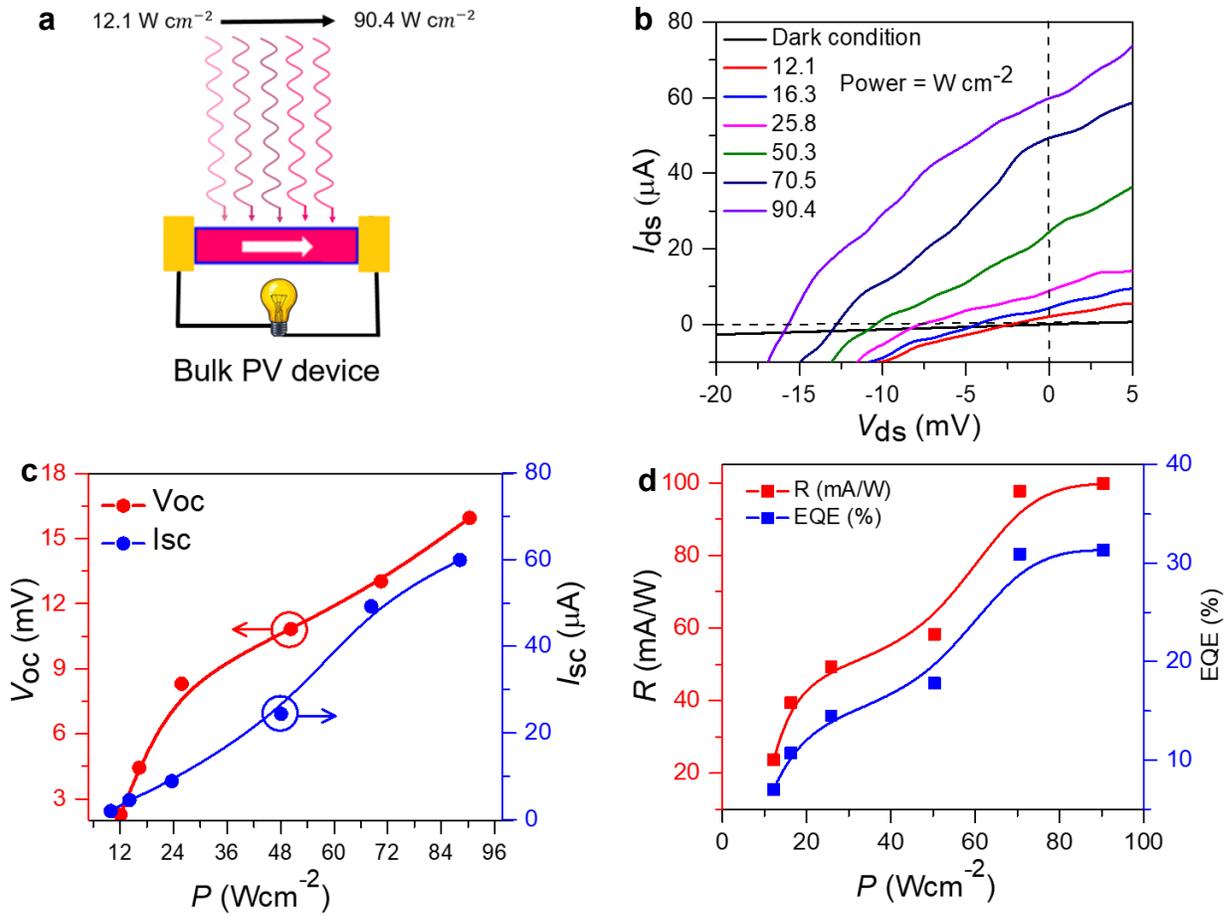

**Figure 6. Electrical characteristics of distorted MoTe$_2$ at different incident laser powers.** (a) Schematic illustrations show bulk PV devices under illumination with different intensities of incident light. (b) I$_{ds}$–V$_{ds}$ characteristics were recorded at different input intensities under laser illumination with a wavelength of 400 nm. (c) Dependence of both Voc and Isc on incident laser power. (d) Dependence of R and EQE on incident laser power.
29